\documentclass[singlecolumn]{revtex4}
\usepackage[pdftex]{graphicx}
\usepackage{amsmath, amssymb} 
\usepackage{amsthm} 
\usepackage{type1cm}
\usepackage{braket}
\usepackage{graphics}
\usepackage{algorithmic} 
\usepackage[version=3]{mhchem}
\usepackage{comment}
\usepackage{bm}
\usepackage{here}

\begin{document}
\title{Modification of the effective action approach for the Leggett mode}
\author{Keita Arimitsu}
\email{ariman55@g.ecc.u-tokyo.ac.jp}
\affiliation{Theoretics Physik, Eidgen\"{o}ssische Technische Hochschule, 8093 Z\"{u}rich, Switzerland}
\altaffiliation{On leave from Department of integrated science, Faculty of Arts and Sciences, the University of Tokyo} 

\begin{abstract}
In multiband superconductors there exists collective excitations which correspond to relative phase fluctuation of each band called the Leggett mode. This is a consequence of the presence of multiple order parameters, which makes multiband systems qualitatively different from singleband systems. Theoretically, this mode can be obtained from the effective action for the phase\cite{sharapov2002effective}. However, the procedure to get the effective action is not clear when one considers relationship between phase of fermion field and that of gap.

 In this paper, the modified procedure to get the effective action is discussed. Careful observation of phase of electrons leads to the conclusion that Hubbard-Stratonovich field should be pair wavefunction, instead of gap. The modified theory is valid for arbitrary strong interband coupling while the earlier approach only deals with weak interband coupling. This paper also discusses experimental observation of the Leggett mode: Raman spectroscopy. Spectral peak which corresponds to the Leggett mode is calculated in the earlier and the modified theory, which shows that they are much different. 

\end{abstract}
\maketitle


\section{Introduction}
Research for collective excitations of superconductors has so long a history as theory of superconductivity itself. One of the most distinctive feature of collective excitations of superconductors comes from the complexity of order parameters: there exist ``phase fluctuation" and ``amplitude fluctuation". The former is called Goldstone mode and the latter is called Higgs mode. This feature is shared by the Higgs mechanism\cite{higgs1964broken} in electroweak interaction, which makes vector bosons massive.

The original BCS theory\cite{bardeen1957theory} assumes that the system is a singleband. Extension to multiband systems was done shortly after the original BCS theory\cite{suhl1959bardeen, moskalenko1959superconductivity}. Multiband superconductors are qualitatively different from singleband counterparts aside from that in multiband systems more than one gap are defined\cite{lin2014ground}. In 1966 Leggett suggested that relative phase fluctuation gives another collective excitation, the Leggett mode\cite{leggett1966number}. The Leggett mode exists only if more than one gap are defined in the system. Thus, there is no counterpart in singleband systems. While Goldstone mode itself is massless, the Leggett mode is generally massive, i.e., dispersion of the Leggett mode takes the form of $\omega_{\mathrm{LG}}^2=m_{\mathrm{LG}}^2+v_{\mathrm{LG}}^2\bm{p}^2$ where $\omega_{\mathrm{LG}}$ is energy dispersion of the Leggett mode and $\bm{p}$ is momentum. This is because in multiband systems generally relative phase is fixed and even homogeneous phase twist which only depends on each band changes energy, though in singleband systems homogeneous phase twist does not change energy.

Multiband superconductors have attracted not only theorists, but also experimentalists particularly after an advent of \ce{MgB2}. It is shown that \ce{MgB2} has three dimensional band originated from $\sigma$ bonding and quasi-two dimensional band from $\pi$ bonding and it shows superconductivity in both bands\cite{ponomarev2004evidence}. Moreover \ce{FeSe} and \ce{Sr2RuO4} are also suggested to be multiband superconductors\cite{ponomarev2011andreev, singh1995relationship}.

The existence of the Leggett mode can be confirmed by the Raman spectroscopy or the tunneling spectroscopy. It has been believed that signatures of the Leggett mode were found in \ce{MgB2} by both experimental methods\cite{blumberg2007observation, ponomarev2004evidence}. Main difficulty of detection of the Leggett mode lies in the fact that estimated $m_{\mathrm{LG}}$ is greater than $2\Delta$. This means that the peak corresponding to the Leggett mode should be placed in quasiparticle continuum. Therefore the Landau damping makes the peak broadened and difficult to discern. 


The Leggett mode can be obtained by path integral formalism which was suggested by Sharapov et al.\cite{sharapov2002effective}. They claimed that this formalism can deal with an arbitrary interband coupling, although Leggett originally obtained this mode by treating interband coupling as a perturbation. The procedure by Sharapov et.al., mimics the procedure to get Goldstone mode. A number of theoretical approaches described above are based on this method\cite{koyama2014collective, karakozov2010theoretical, anishchanka2007collective, mou2015strong, huang2016leggett}. On the other hand, careful inspection into the relation between gap and fermionic field can show that this procedure is not valid particularly when interband coupling is large. This paper provides a consistent effective action approach for the Leggett mode. This paper is organized in the following way: first two band model is analyzed for both a neutral and a charged superconductor. Next the focus is on the Raman spectroscopy, emphasizing on the difference between the present approach and the earlier approach.

\section{Two band model}
As Leggett and Sharapov et al. did, a system with two bands is considered. Extension to systems with arbitrary number of bands is straightforward.

First, a neutral superconductor with two bands is described by the following Hamiltonian ($\hbar=1$):
\begin{equation*}
\begin{split}
H_{\mathrm{neutral}}=&\sum_{i=1,2}\sum_{\sigma=\uparrow, \downarrow}\int d\bm{r}\psi_{i\sigma}^\dagger(\bm{r})\left(-\frac{\nabla^2}{2m_i}-\mu_i\right)\psi_{i\sigma}(\bm{r})-\sum_{ij}V_{ij}\int d\bm{r}\psi_{i\uparrow}^\dagger(\bm{r})\psi_{i\downarrow}^\dagger(\bm{r})\psi_{j\downarrow}(\bm{r})\psi_{j\uparrow}(\bm{r}), 
\end{split}
\end{equation*}
where $i, j$ denote band indexes, $\sigma=\uparrow, \downarrow$ denotes spin and $m_i$ is an effective mass of electrons in $i$th band. For simplicity let us assume $V_{12}=V_{21}=J\in\mathbb{R}$. 

To describe a charged superconductor, one should add Coulomb interaction to $H_{\mathrm{neutral}}$
\begin{equation*}
\begin{split}
H_{\mathrm{charged}}=H_{\mathrm{neutral}}+\frac{1}{2}\int&d\bm{r}_1d\bm{r}_2\rho(\bm{r}_1)V_{\mathrm{C}}(\bm{r}_1-\bm{r}_2)\rho(\bm{r}_2),
\end{split}
\end{equation*}
where 
\begin{equation*}
\begin{split}
\rho(\bm{r})&=\sum_{i\sigma}\psi_{i\sigma}^\dagger(\bm{r})\psi_{i\sigma}(\bm{r})-n,\\
V_{\mathrm{C}}(\bm{r}_1-\bm{r}_2)&=\frac{e^2}{|\bm{r}_1-\bm{r}_2|},
\end{split}
\end{equation*}
with $n$ being background charge.

The partition function is 
\begin{equation*}
\begin{split}
Z&=\int\mathcal{D}\psi^\dagger\mathcal{D}\psi e^{-S},\\
S&=\int_0^\beta d\tau\int d\bm{r}\sum_{i\sigma}\psi_{i\sigma}^\dagger(x)\partial_\tau\psi_{i\sigma}(x)+H(\tau).
\end{split}
\end{equation*}
Here $\psi(x)=\psi(\tau,\bm{r})$ is a Grassman number depending on imaginary time $\tau$ and coordinate $\bm{r}$. 
\subsection{Mean-field theory}
For a neutral superconductor, let us introduce Hubbard-Stratonovich field $\Phi_i(x), \Phi_i^*(x)$ in order to get the action which is quadratic in $\psi^\dagger, \psi$.

With those fields, the action $S_{\mathrm{neutral}}$ becomes
\begin{equation}\label{1}
\begin{split}
&S_{\mathrm{neutral}}=S_0+S_{\mathrm{pair}},\\
&S_0=\int_0^\beta d\tau\int d\bm{r}\sum_{i\sigma}\psi_{i\sigma}^\dagger(x)\left(\partial_\tau-\frac{\nabla^2}{2m_i}-\mu_i\right)\psi_{i\sigma}(x),\\
&S_{\mathrm{pair}}=\int_0^\beta d\tau\int d\bm{r}-\sum_{ij}\left(V_{ij}\Phi_i^*(x)\psi_{j\downarrow}(x)\psi_{j\uparrow}(x)+h.c.\right)+\sum_{ij}V_{ij}\Phi_i^*(x)\Phi_j(x).
\end{split}
\end{equation}
Notice that this Hubbard-Stratonovich transformation is different from that in a conventional manner. Conventionally $S_{\mathrm{pair}}$ is
\begin{equation}\label{2}
\begin{split}
S_{\mathrm{pair}}=&\int_0^\beta d\tau\int d\bm{r}-\sum_i\left(\Delta^*_i(x)\psi_{i\downarrow}(x)\psi_{i\uparrow}(x)+h.c.\right).+\sum_{ij}\left(V^{-1}\right)^{ij}\Delta^*_i(x)\Delta_j(x), 
\end{split}
\end{equation}
where $\left(V^{-1}\right)^{ij}$ is an inverse matrix of $V_{ij}$ such that 
\begin{equation*}
\sum_jV_{ij}\left(V^{-1}\right)^{jk}=\delta_{ik}.
\end{equation*}
Relation between Hubbard-Stratonovich fields in (\ref{1}) and (\ref{2}) is
\begin{equation}\label{3}
\Delta_i(x)=\sum_jV_{ij}\Phi_j(x).
\end{equation}
From (\ref{3}) it is clear that $\Phi_i(x)$ represents the pair wavefunction.

For a charged superconductor one should introduce another Hubbard-Stratonovich field $\varphi(x)$ as follows:
\begin{equation*}
\begin{split}
S_{\mathrm{charged}}&=S_{\mathrm{neutral}}+S_{\mathrm{C}}\\
S_{\mathrm{C}}&=-i\int_0^\beta d\tau\int d\bm{r}\rho(x)\varphi(x)-\frac{1}{8\pi e^2}\int_0^\beta d\tau\int d\bm{r}\varphi(x)\nabla^2\varphi(x).
\end{split}
\end{equation*}
Note that the second term depends on the spatial dimension of the system and here we have assumed three dimensional system. 

For the moment, a neutral superconductor will be considered.  With Nambu spinor $\Psi_i(x)=\begin{pmatrix} \psi_{i\uparrow}(x) & \psi_{i\downarrow}^\dagger(x) \end{pmatrix}^{\mathrm{T}}$, the action (\ref{1}) is made quadratic
\begin{equation*}
\begin{split}
S_{\mathrm{neutral}}=&\int_0^\beta d\tau\int d\bm{r}\sum_i\Psi_i^\dagger(x)\left(-G_{0i}^{-1}\right)\Psi_i(x)+\sum_{ij}V_{ij}\Phi_i^*(x)\Phi_j(x)\\
G^{-1}_{0i}=&\begin{pmatrix}-\partial_\tau+\frac{\nabla^2}{2m_i}+\mu_i & \sum_jV_{ij}\Phi_j(x)\\ \sum_jV_{ij}\Phi_j^*(x) & -\partial_\tau-\frac{\nabla^2}{2m_i}-\mu_i\end{pmatrix}.
\end{split}
\end{equation*}
The first term can be integrated out. After integration, one gets
\begin{equation}\label{4}
\begin{split}
&S_{\mathrm{neutral}}=\int_0^\beta d\tau\int d\bm{r}\sum_{ij}V_{ij}\Phi_i^*(x)\Phi_j(x)-\sum_i\mathrm{LnDet}G_{0i}^{-1}.
\end{split}
\end{equation}
The value $\Phi_i(x)$ is obtained by minimizing the action with respect to $\Phi_i^*(x)$, i.e.,
\begin{equation}\label{gapeq}
\frac{\delta S_{\mathrm{neutral}}}{\delta\Phi_i^*(x)}=0.
\end{equation}
In mean-field approximation, it is assumed that $\Phi_i(x)$ is independent of $x$, i.e., $\Phi_i(x)=\Phi_i$. Combination of (\ref{gapeq}) and this approximation gives the gap equation
\begin{equation}\label{6}
\sum_jV_{ij}\Phi_j=\sum_{jj'}V_{ij}\int\frac{d\bm{k}}{(2\pi)^d}\frac{V_{jj'}\Phi_{j'}}{2E_{j}(\bm{k})}\tanh\left(\frac{\beta E_{j}(\bm{k})}{2}\right), 
\end{equation}
where $E_{i}(\bm{k})=\sqrt{\xi_{i}(\bm{k})^2+\left|\sum_jV_{ij}\Phi_j\right|^2}$ with $\xi_{i}(\bm{k})=\frac{\hbar^2\bm{k}^2}{2m_i}-\mu_{i}$ and $d$ is the dimension of the system. In fact, by using the relation (\ref{3}), (\ref{6}) can be written as
\begin{equation*}
\Delta_i=\sum_jV_{ij}\int\frac{d\bm{k}}{(2\pi)^d}\frac{\Delta_j}{2E_j(\bm{k})}\tanh\left(\frac{\beta E_j(\bm{k})}{2}\right),
\end{equation*}
where $d$ is spatial dimension of the given system. Note that this integral appears to be divergent. Therefore prescription should be set to remove this divergence. One way is to assume that interaction is BCS type. In this case, $k$ integral is limited by modifying interaction,
\begin{equation*}
V_{ij}\rightarrow V_{ij}\times\Theta\left(\omega_D-|\xi_j(\bm{k})|\right),
\end{equation*}
where $\Theta$ is a step function and $\omega_D$ is the Debye frequency. Another way is to assume a lattice system. In this case, cutoff is as follows:
\begin{equation*}
\int\frac{d\bm{k}}{(2\pi)^d}\rightarrow\int_{\bm{k}\in\mathrm{B.Z.}}\frac{d\bm{k}}{(2\pi)^d},
\end{equation*}
where B.Z. denotes Brillouin zone.

In this paper, we assume that divergence is removed by some prescription, but do not assume any specific prescription. The following argument holds for any prescription as long as it is introduced.

In what follows $\Phi_i\in\mathbb{R}$ is assumed for simplicity. This assumption is always possible for two band systems.

\subsection{Phase fluctuation}
The Hubbard-Stratonovich field $\Phi_i(x)$ is a complex scalar field. Thus the phase of $\Phi_i(x)$ can fluctuate around the mean-field value and this leads to the Leggett mode and BAG (Bogoliubov-Anderson-Goldstone) mode. 

To get an effective action for the phase, let us consider the following set of transformation:
\begin{equation}\label{7}
\begin{split}
\psi_{i\sigma}(x)&\rightarrow e^{i\theta_i(x)/2}\psi_{i\sigma}(x)\\
\Phi_i(x)&\rightarrow\Phi_i e^{i\theta_i(x)},
\end{split}
\end{equation}
where $\Phi_i$ is a mean-field value of $\Phi_i(x)$.


Generally speaking, one can arbitrarily introduce phase fluctuations. Therefore one must look for consistent relationship among fluctuations.


In canonical quantization formalism pair wavefunction is introduced as follows:
\begin{equation*}
\begin{split}
\Phi_i&=\braket{\psi_{i\downarrow}(\bm{r})\psi_{i\uparrow}(\bm{r})}\\
\Phi_i^*&=\braket{\psi_{i\uparrow}^\dagger(\bm{r})\psi_{i\downarrow}^\dagger(\bm{r})}.
\end{split}
\end{equation*}
Therefore in path integral formalism it is natural to require the following relation between fermionic fields $\psi_{i\sigma}(x), \psi_{i\sigma}^\dagger(x)$ and Hubbard-Stratonovich fields $\Phi_i(x), \Phi_i^*(x)$:
\begin{equation}\label{8}
\begin{split}
\Phi_i(x)&=\psi_{i\downarrow}(x)\psi_{i\uparrow}(x),\\
\Phi_i^*(x)&=\psi_{i\uparrow}^\dagger(x)\psi_{i\downarrow}^\dagger(x).
\end{split}
\end{equation}
This relation implies that phase fluctuations should satisfy
\begin{equation}\label{9}
\Phi_i(x)\rightarrow\Phi'_i(x)=\psi_{i\downarrow}(x)\psi_{i\uparrow}(x)e^{i\theta_i(x)}=\Phi_ie^{i\theta_i(x)}.
\end{equation}
(\ref{9}) is nothing but (\ref{7}). Therefore, the set of phase fluctuation in (\ref{7}) is consistent.

In the earlier approach, the following phase fluctuation was considered:
\begin{equation}\label{10}
\begin{split}
\Delta_i(x)&\rightarrow e^{i\theta_i(x)}\Delta_i\\
\Delta_i^*(x)&\rightarrow e^{-i\theta_i(x)}\Delta_i^*,
\end{split}
\end{equation}
where $\Delta_i, \Delta_i^*$ are the mean field value of $i$th band. In the canonical quantization formalism, the gap is defined such as
\begin{equation*}
\begin{split}
\Delta_i&=\sum_jV_{ij}\braket{\psi_{j\downarrow}(\bm{r})\psi_{j\uparrow}(\bm{r})}\\
\Delta_i^*&=\sum_jV_{ij}\braket{\psi_{j\uparrow}^\dagger(\bm{r})\psi_{j\downarrow}^\dagger(\bm{r})}.
\end{split}
\end{equation*}
From this relation it is also natural to require the following:
\begin{equation*}
\begin{split}
\Delta_i(x)&=\sum_jV_{ij}\psi_{j\downarrow}(x)\psi_{j\uparrow}(x)=\sum_jV_{ij}\Phi_j(x)\\
\Delta_i^*(x)&=\sum_jV_{ij}\psi_{j\uparrow}^\dagger(x)\psi_{j\downarrow}^\dagger(x)=\sum_jV_{ij}\Phi_j^*(x),
\end{split}
\end{equation*}
where relation (\ref{7}) is used. If the phase of fermionic field is twisted $\psi_{i\sigma}(x)\rightarrow e^{i\theta_i(x)/2}\psi_{i\sigma}(x)$, this leads to 
\begin{equation}\label{11}
\begin{split}
\Delta_i(x)&\rightarrow\Delta_i'(x)=\sum_jV_{ij}\Phi_je^{i\theta_j(x)}\\
\Delta_i^*(x)&\rightarrow\Delta_i'^*(x)=\sum_jV_{ij}\Phi_j^*e^{-i\theta_j(x)}.
\end{split}
\end{equation}
Generally (\ref{11}) is not equivalent to (\ref{10}). Therefore to achieve (\ref{10}) fields should be transformed such as
\begin{equation*}
\begin{split}
\Delta_i(x)&\rightarrow\Delta_i'(x)=\sum_{ij}V_{ij}\left(\Phi_j+\delta\Phi_j(x)\right)e^{i\theta_j(x)}\\
\Delta_i^*(x)&\rightarrow\Delta_i'^*(x)=\sum_{ij}V_{ij}\left(\Phi_j^*+\delta\Phi_j^*(x)\right)e^{-i\theta_j(x)},
\end{split}
\end{equation*}
with constraints
\begin{equation*}
\begin{split}
\Delta_i(x)e^{i\theta_i(x)}&=\left(\sum_jV_{ij}\Phi_j\right)e^{i\theta_i(x)}=\sum_{ij}V_{ij}\left(\Phi_j+\delta\Phi_j(x)\right)e^{i\theta_j(x)}\\
\Delta_i^*(x)e^{-i\theta_i(x)}&=\left(\sum_jV_{ij}\Phi^*_j\right)e^{-i\theta_i(x)}=\sum_{ij}V_{ij}\left(\Phi_j+\delta\Phi^*_j(x)\right)e^{-i\theta_j(x)},
\end{split}
\end{equation*}
where $\delta\Phi(x), \delta\Phi^*(x)$ are so called ``amplitude fluctuation", Higgs mode. Since Higgs mode is generally massive, there is no reason to believe this way of fluctuation gives low energy excitation. Indeed, it will be shown that collective excitation defined in (\ref{7}) gives lower energy than that defined in (\ref{10}).  

From this observation it can be concluded that in order to obtain the effective action of the phase in two-band system, one should take pair wavefunction as a Hubbard-Stratonovich field.

\subsection{The effective action of the phase}
\subsubsection{Collective excitations in a neutral superconductor}\label{222}
By the set of phase fluctuations defined in (\ref{7}), the action becomes the following form:
\begin{equation*}
\begin{split}
S_{\mathrm{neutral}}&=S_0+S_{\mathrm{pair}}\rightarrow S_{\mathrm{neutral}}'=S_0'+S'_{\mathrm{pair}}\\
S_0'&=S_0+\int_0^\beta d\tau\int d\bm{r}\sum_i\Psi_i^\dagger(x)\Sigma_0^i\Psi_i(x)\\
S_{\mathrm{pair}}'&=S_{\mathrm{pair}}+\int_0^\beta d\tau\int d\bm{r}\sum_i\Psi_i^\dagger(x)\Sigma_{\mathrm{pair}}^i\Psi_i(x)-\frac{1}{2}\int_0^\beta d\tau\int d\bm{r}\sum_{ij}\Phi_i\Phi_j\left(\theta_i(x)-\theta_j(x)\right)^2\\
\Sigma_0^i&=-\frac{i}{2m}\left(\nabla\theta_i(x)\cdot\nabla+\frac{1}{2}\nabla^2\theta_i(x)\right)\tau^0+\left(\frac{i}{2}\partial_\tau\theta_i(x)+\frac{1}{8m_i}\left(\nabla\theta_i(x)\right)^2\right)\tau^3\\
\Sigma_{\mathrm{pair}}^i&=-J\Phi_{\overline{i}}\left[\left(\theta_i(x)-\theta_{\overline{i}}(x)\right)\tau^2-\frac{1}{2}\left(\theta_i(x)-\theta_{\overline{i}}(x)\right)^2\tau^1\right],
\end{split}
\end{equation*}
where $\overline{i}$ denotes the other band than $i$th band, i.e., $\overline{1}=2, \overline{2}=1$, $\tau^i (i=1,2,3)$ is $i$th component of Pauli matrix, and $\tau^0$ is an unit matrix.
For deriving $S'_{\mathrm{pair}}$, exponential is expanded up to second order in $\theta$, i.e., 
\begin{equation*}
e^{-i(\theta_i(x)-\theta_j(x))}\cong1-i(\theta_i(x)-\theta_j(x))-\frac{1}{2}(\theta_i(x)-\theta_j(x))^2.
\end{equation*}

$\Psi, \Psi^\dagger$ can be integrated out by the following identity:
\begin{equation*}
\mathrm{LnDet}\left(G_{0i}^{-1}-\Sigma^i\right)=\mathrm{LnDet}G_{0i}^{-1}-\sum_{n=1}^\infty\frac{1}{n}\mathrm{Tr}\left(G_{0i}\Sigma^i\right)^n,
\end{equation*}
where $\Sigma^i=\Sigma_0^i+\Sigma_{\mathrm{pair}}^i$ and
\begin{equation*}
G_{0i}(i\omega_n,\bm{k})=\frac{-1}{\omega_n^2+E_i(\bm{k})^2}\begin{pmatrix}i\omega_n+\xi_i(\bm{k}) & -\Delta_i \\ -\Delta_i & i\omega_n-\xi_i(\bm{k})\end{pmatrix},
\end{equation*}
with $\omega_n=(2n+1)\pi\beta^{-1}$ being Matsubara frequency for fermionic fields and $\Delta_i$ being defined by (\ref{4}).

The second part of this identity gives the effective action for the phase.
\begin{equation*}
\begin{split}
\sum_{n=1}^\infty\frac{1}{n}\mathrm{Tr}\left(G_{0i}\Sigma^i\right)^n=&\sum_{l=-\infty}^\infty\int\frac{d\bm{p}}{(2\pi)^d}\begin{pmatrix} \theta_1(-p) & \theta_2(-p)\end{pmatrix}\mathcal{M}_i
\begin{pmatrix} \theta_1(p) \\ \theta_2(p) \end{pmatrix}\\& +O(\theta^3),
\end{split}
\end{equation*}
where $\theta_i(p)=\theta_i(i\nu_l,\bm{p})$ with $\nu_l=2l\pi\beta^{-1}$ being Matsubara frequency for bosonic field. $\mathcal{M}_i$ is a $2\times2$ matrix with each component being
\begin{equation*}
\begin{split}
\left(\mathcal{M}_i\right)_{ii}&=\frac{1}{2\beta}\sum_n\int\frac{d\bm{k}}{(2\pi)^d}\left[-\frac{\nu_l^2}{4}\pi^{33}_i(k,p)-\frac{i\nu_l}{8m_i}(2\bm{k}+\bm{p})\cdot\bm{p}\left(\pi^{30}_i(k,p)+\pi^{03}_i(k,p)\right)+\frac{1}{(4m_i)^2}\left((2\bm{k}+\bm{p})\cdot\bm{p}\right)^2\pi^{00}_i(k,p)\right.\\&\left.-\frac{iJ\Phi_{\overline{i}}}{4m}(2\bm{k}+\bm{p})\cdot\bm{p}\left(\pi^{20}_i(k,p)-\pi^{02}_i(k,p)\right)+J\Phi_{\overline{i}}\frac{\nu_l}{2}\left(\pi^{32}_i(k,p)-\pi^{23}(k,p)\right)+J^2\Phi_{\overline{i}}^2\pi^{22}_i(k,p)\right]+\frac{n_i}{8m_i}\bm{p}^2\\
\left(\mathcal{M}_i\right)_{i\overline{i}}&=\frac{-1}{2\beta}\sum_n\int\frac{d\bm{k}}{(2\pi)^d}\left[\frac{iJ\Phi_{\overline{i}}}{4m_i}(2\bm{k}+\bm{p})\cdot\bm{p}\pi^{02}_i(k,p)+J\Phi_{\overline{i}}\frac{\nu_l}{2}\pi^{32}_i(k,p)+J^2\Phi_{\overline{i}}^2\pi^{22}_i(k,p)\right]\\
\left(\mathcal{M}_i\right)_{\overline{i}i}&=\frac{1}{2\beta}\sum_n\int\frac{d\bm{k}}{(2\pi)^d}\left[\frac{iJ\Phi_{\overline{i}}}{4m_i}(2\bm{k}+\bm{p})\cdot\bm{p}\pi_i^{20}(k,p)+J\Phi_{\overline{i}}\frac{\nu_l}{2}\pi_i^{23}(k,p)-J^2\Phi_{\overline{i}}^2\pi_i^{22}(k,p)\right]\\
\left(\mathcal{M}_i\right)_{\overline{i}\overline{i}}&=\frac{1}{2\beta}\sum_n\int\frac{d\bm{k}}{(2\pi)^d}J^2\Phi_{\overline{i}}^2\pi^{22}_i(k,p),
\end{split}
\end{equation*}
where
\begin{equation*}
\begin{split}
\pi^{\mu\nu}_i(k,p)&=\mathrm{tr}\left[G_{0i}(\bm{k},i\omega_n)\tau^\mu G_{0i}(\bm{k}+\bm{p},i\omega_n+i\nu_l)\tau^\nu\right],
\end{split}
\end{equation*}
and $n_i$ is density of electron of $i$th band. Note that to obtain the result above, as for terms which are quadratic in $\theta_i(x)$ it suffices to substitute the mean-field value with Nambu spinors (or Fermion field), i.e., 
\begin{equation*}
\begin{split}
&\Psi_i^\dagger(x)\left(\frac{1}{8m_i}\left(\nabla\theta_i(x)\right)^2\tau^3\right)\Psi_i(x)=\left(\sum_{\sigma}\psi_{i\sigma}^\dagger(x)\psi_{i\sigma}(x)\right)\frac{1}{8m_i}\left(\nabla\theta_i(x)\right)^2\\
&\rightarrow\left(\sum_{\sigma}\braket{\psi_{i\sigma}^\dagger(x)\psi_{i\sigma}(x)}\right)\frac{1}{8m_i}\left(\nabla\theta_i(x)\right)^2
=\frac{n_i}{8m_i}\left(\nabla\theta_i(x)\right)^2,
\end{split}
\end{equation*}
and
\begin{equation*}
\begin{split}
&\Psi_i^\dagger(x)\left(\frac{1}{2}J\Phi_{\overline{i}}\left(\theta_i(x)-\theta_{\overline{i}}(x)\right)^2\tau^1\right)\Psi_i(x)=\frac{1}{2}J\Phi_{\overline{i}}\left(\psi_{i\downarrow}(x)\psi_{i\uparrow}(x)+\psi_{i\uparrow}^\dagger(x)\psi_{i\downarrow}^\dagger(x)\right)\left(\theta_i(x)-\theta_{\overline{i}}(x)\right)^2\\
&\rightarrow\frac{1}{2}J\Phi_{\overline{i}}\left(\braket{\psi_{i\downarrow}(x)\psi_{i\uparrow}(x)}+\braket{\psi_{i\uparrow}^\dagger(x)\psi_{i\downarrow}^\dagger(x)}\right)\left(\theta_i(x)-\theta_{\overline{i}}(x)\right)^2=J\Phi_i\Phi_{\overline{i}}\left(\theta_i(x)-\theta_{\overline{i}}(x)\right)^2.
\end{split}
\end{equation*}
Finally the action is
\begin{equation*}
\begin{split}
S'_{\mathrm{neutral}}=&\sum_{l=-\infty}^\infty\int\frac{d\bm{p}}{(2\pi)^d}\begin{pmatrix} \theta_1(-p) & \theta_2(-p) \end{pmatrix}G_{\theta;\mathrm{neutral}}^{-1}\begin{pmatrix} \theta_1(p) \\ \theta_2(p)\end{pmatrix}+\beta\Omega\sum_{ij}V_{ij}\Phi_i^*\Phi_j,
\end{split}
\end{equation*}
where $G_{\theta;\mathrm{neutral}}^{-1}$ is a $2\times2$ matrix with components being
\begin{equation*}
\begin{split}
G_{\theta;\mathrm{neutral}}^{-1}=\left(\mathcal{M}_1+\mathcal{M}_2\right)+J\Phi_1\Phi_2\left(\tau^0-\tau^1\right)
\end{split}
\end{equation*}
Energy dispersions are obtained by $\mathrm{det}G_{\theta;\mathrm{neutral}}^{-1}=0$ followed by analytic continuation $i\nu_l\rightarrow\omega+i0$.

Let us assume the spatial dimension of the given system is three, i.e., $d=3$. At zero temperature and at hydrodynamic limit, i.e., $\nu_l\rightarrow0, \bm{p}\rightarrow\bm{0}$ by substituting $\pi^{\mu\nu}_i(k,p)\cong\pi^{\mu\nu}_i(k,0)$ each component of $G_{\theta;\mathrm{neutral}}$ is as follows:
\begin{equation}\label{12}
\begin{split}
\left(G_{\theta;\mathrm{neutral}}^{-1}\right)_{11}&=\frac{1}{4}\left(\rho_1\nu_l^2+\rho_1c_1^2\bm{p}^2+4\mu_{\mathrm{LG}}\right)\\
\left(G_{\theta;\mathrm{neutral}}^{-1}\right)_{12}&=-\mu_{\mathrm{LG}}\\
\left(G_{\theta;\mathrm{neutral}}^{-1}\right)_{21}&=-\mu_{\mathrm{LG}}\\
\left(G_{\theta;\mathrm{neutral}}^{-1}\right)_{22}&=\frac{1}{4}\left(\rho_2\nu_l^2+\rho_2c_2^2\bm{p}^2+4\mu_{\mathrm{LG}}\right)\\
\mu_{\mathrm{LG}}&=\sum_i\lambda_i\left(\frac{1}{2}-\lambda_i\right)\Phi_i\Delta_i,
\end{split}
\end{equation}
where 
\begin{equation}\label{13}
\lambda_i=\frac{J\Phi_{\overline{i}}}{\Delta_i}, 
c_i^2=\frac{v_{Fi}^2}{3},
\end{equation}
with $\rho_i$ being the density of state of electrons in $i$th band and $v_{Fi}$ being Fermi velocity of electrons in $i$th band. Note that the following relations are used for the derivation of (\ref{12}):
\begin{equation}\label{gapeq}
\begin{split}
\Delta_i&=V_{ii}\Phi_i+J\Phi_{\overline{i}}\quad(\text{the gap equation})\\
\Phi_i&=\int\frac{d\bm{k}}{(2\pi)^d}\frac{\Delta_i}{2E_i(\bm{k})}\tanh\left(\frac{\beta E_i(\bm{k})}{2}\right).
\end{split}
\end{equation}

From (\ref{12}) dispersions of BAG mode and the Leggett mode are the following:
\begin{equation}\label{14}
\begin{split}
\omega_{\mathrm{BAG}}^2&=c^2\bm{p}^2\\
\left(\omega_{\mathrm{LG}}^{\mathrm{neutral}}\right)^2&=\left(\omega_0^{\mathrm{modified}}\right)^2+v^2\bm{p}^2,
\end{split}
\end{equation}
where 
\begin{equation*}
\begin{split}
c^2&=\frac{\rho_1c_1^2+\rho_2c_2^2}{\rho_1+\rho_2}\\
\left(\omega_0^{\mathrm{modified}}\right)^2&=4\frac{\rho_1+\rho_2}{\rho_1\rho_2}\mu_{\mathrm{LG}}, 
v^2=\frac{\rho_1c_2^2+\rho_2c_1^2}{\rho_1+\rho_2}.
\end{split}
\end{equation*}
Within this approximation BAG mode does not change while the Leggett mode is modified.

Compare this result with the following derived by Leggett in his original paper or Sharapov et al.:
\begin{equation}\label{15}
\begin{split}
\left(\omega_{\mathrm{LG}}^{\mathrm{initial}}\right)^2&=\left(\omega_0^{\mathrm{initial}}\right)^2+v^2\bm{p}^2,
\end{split}
\end{equation}
where 
\begin{equation*}
\left(\omega_0^{\mathrm{initial}}\right)^2=4\frac{\rho_1+\rho_2}{\rho_1\rho_2}\frac{J\Delta_1\Delta_2}{V_{11}V_{22}-J^2}.
\end{equation*}

$\omega_0^{\mathrm{initial}}$ can be expressed by $\lambda_i$ as follows:
\begin{equation}\label{lambda}
\left(\omega_0^{\mathrm{initial}}\right)^2=\frac{1}{2}\sum_i\frac{\lambda_i\Phi_i\Phi_{\overline{i}}}{(1-\lambda_i)(1-\lambda_{\overline{i}})\Phi_{\overline{i}}\Delta_i\Delta_{\overline{i}}-\lambda_i^2\Phi_i\Delta_i^2}\Delta_i^2\Delta_{\overline{i}}^2,
\end{equation}
where (\ref{lambda}) is derived by using the definition (\ref{13}) and (\ref{gapeq}).

One can expand $\omega_0^{\mathrm{modified}}$ and $\omega_0^{\mathrm{initial}}$ with respect to $\lambda_i$ to find that both are of form

\begin{equation}\label{linear}
\omega_0^2=\frac{1}{2}\sum_i\lambda_i\Phi_i\Delta_i+O(\lambda_i^2).
\end{equation}

From (\ref{linear}) it can be concluded that up to linear order in $\lambda_i$ implying small interband coupling, results from both theories are the same.
\subsubsection{Collective excitations in a charged superconductor}
In what follows we assume $d=3$.

In order to consider collective excitations in a charged superconductor, one should add $S_c$ to $S_{\mathrm{neutral}}$. Since $S_c$ does not change under the transformation (\ref{7}), the action becomes the following form:
\begin{equation*}
\begin{split}
S_{\mathrm{charged}}&=S_{\mathrm{neutral}}+S_c\rightarrow S'_{\mathrm{charged}}=S'_{\mathrm{neutral}}+S_c\\
S_c&=-\frac{1}{8\pi e^2}\int_0^\beta d\tau\int d\bm{r}\varphi(x)\nabla^2\varphi(x)-i\int_0^\beta d\tau\int d\bm{r}\sum_i\varphi(x)\Psi_i^\dagger(x)\tau^3\Psi_i(x).
\end{split}
\end{equation*}
With the same procedure as we saw to get the effective action for the phase in the previous section, the effective action for phase in a charged superconductor is the following:
\begin{equation*}
\begin{split}
S_{\mathrm{charged}}'&=\sum_{l=-\infty}^\infty\int\frac{d\bm{p}}{(2\pi)^3}\begin{pmatrix} \theta_1(-p)&\theta_2(-p)&\varphi(-p)\end{pmatrix}G_{\theta;\mathrm{charged}}^{-1}\begin{pmatrix} \theta_1(p) \\ \theta_2(p) \\ \varphi(p) \end{pmatrix}+\beta\Omega\sum_{ij}V_{ij}\Phi_i\Phi_j, 
\end{split}
\end{equation*}
where $\varphi(p)=\varphi(i\nu_l,\bm{p})$ is Fourier transform of the Hubbard-Stratonovich field $\varphi(x)$ and $G_{\theta;\mathrm{charged}}^{-1}$ is a $3\times3$ matrix with each component being
\begin{equation*}
\begin{split}
\left(G_{\theta;\mathrm{charged}}^{-1}\right)_{13}&=\frac{1}{2\beta}\sum_n\int\frac{d\bm{k}}{(2\pi)^3}\frac{i\nu_l}{2}\pi^{33}_1(k,p)-\frac{1}{4m_1}(2\bm{k}+\bm{p})\cdot\bm{p}\pi_1^{03}(k,p)+iJ\Phi_2\pi^{23}_1(k,p)-iJ\Phi_1\pi^{23}_2(k,p)\\
\left(G_{\theta;\mathrm{charged}}^{-1}\right)_{23}&=\frac{1}{2\beta}\sum_n\int\frac{d\bm{k}}{(2\pi)^3}\frac{i\nu_l}{2}\pi^{33}_2(k,p)-\frac{1}{4m_2}(2\bm{k}+\bm{p})\cdot\bm{p}\pi_2^{03}(k,p)+iJ\Phi_1\pi^{23}_2(k,p)-iJ\Phi_2\pi^{23}_1(k,p)\\
\left(G_{\theta;\mathrm{charged}}^{-1}\right)_{31}&=\left(\left(G_{\theta;\mathrm{charged}}^{-1}\right)_{13}\right)^*,
\left(G_{\theta;\mathrm{charged}}^{-1}\right)_{32}=\left(\left(G_{\theta;\mathrm{charged}}^{-1}\right)_{23}\right)^*\\
\left(G_{\theta;\mathrm{charged}}^{-1}\right)_{33}&=\frac{1}{2\beta}\sum_n\int\frac{d\bm{k}}{(2\pi)^3}-\left(\pi^{33}_1(k,p)+\pi^{33}_2(k,p)\right)+\frac{\bm{p}^2}{8\pi e^2}\\
\left(G_{\theta;\mathrm{charged}}^{-1}\right)_{mn}&=\left(G_{\theta;\mathrm{neutral}}^{-1}\right)_{mn}\quad(m,n=1,2).
\end{split}
\end{equation*}
$\mathrm{det}G_{\theta;\mathrm{charged}}^{-1}=0$ followed by analytic continuation $i\nu_l\rightarrow\omega+i0$ gives dispersions. 

At zero temperature and in the hydrodynamic limit, i.e., $\pi^{\mu\nu}_i(k,p)\cong\pi^{\mu\nu}_i(k,0)$, each component becomes
\begin{equation*}
\begin{split}
\left(G_{\theta;\mathrm{charged}}^{-1}\right)_{13}&=-\frac{1}{2}\rho_1i\nu_l, \left(G_{\theta;\mathrm{charged}}^{-1}\right)_{23}=-\frac{1}{2}\rho_2i\nu_l,\\
\end{split}
\end{equation*}
\begin{equation*}
\begin{split}
\left(G_{\theta;\mathrm{charged}}^{-1}\right)_{33}&=\rho_1+\rho_2+\frac{\bm{p}^2}{8\pi e^2}.
\end{split}
\end{equation*}
In the limit $\bm{p}^2/8\pi e^2\rightarrow0$ where the bosonic field $\varphi(x)$ can be regarded as potential term, the Leggett mode becomes
\begin{equation}\label{16}
\begin{split}
\left(\omega_{\mathrm{LG}}^{\mathrm{charged}}\right)^2&=\left(\omega_0^{\mathrm{modified}}\right)^2+\left(v^{\mathrm{charged}}\right)^2\bm{p}^2,
\end{split}
\end{equation}
where 
\begin{equation*}
\left(v^{\mathrm{charged}}\right)^2=\frac{(\rho_1+\rho_2)c_1^2c_2^2}{\rho_1c_1^2+\rho_2c_2^2}.
\end{equation*}

Due to the presence of Coulomb interaction, BAG mode acquires mass, which can be obtained by neglecting the interband interaction $J=0$:
\begin{equation}\label{17}
\begin{split}
\left(\omega_{\mathrm{BAG}}^{\mathrm{charged}}\right)^2&=\left(\omega_{\mathrm{plasma}}\right)^2+\left(c^{\mathrm{charged}}\right)^2\bm{p}^2,
\end{split}
\end{equation}
where
\begin{equation*}
\begin{split}
\left(\omega_{\mathrm{plasma}}\right)^2&=8\pi e^2(\rho_1c_1^2+\rho_2c_2^2)\\ \left(c^{\mathrm{charged}}\right)^2&=\frac{\rho_1c_1^4+\rho_2c_2^4}{\rho_1c_1^2+\rho_2c_2^2}.
\end{split}
\end{equation*}
Note that to derive results (\ref{16}) and (\ref{17}), it is assumed that $\omega_{\mathrm{LG}}^{\mathrm{charged}}$ is sufficiently smaller than $\omega_{\mathrm{plasma}}$, i.e., $\omega_{\mathrm{LG}}^{\mathrm{charged}}\ll\omega_{\mathrm{plasma}}$.

As is the case with a neutral superconductor, ignoring more than linear term with respect to $\lambda_i$ gives the same result with Sharapov et al. such that
\begin{equation}\label{18}
\begin{split}
\left(\omega_{\mathrm{LG}}^{\mathrm{charged}}\right)^2&=\left(\omega_0^{\mathrm{initial}}\right)^2+\left(v^{\mathrm{charged}}\right)^2\bm{p}^2,
\end{split}
\end{equation}
up to linear order in $J$. 
%
\section{Vanishing of the Leggett mode}
In this section, $V_{11}, V_{22}, J$ are all real and positive for simplicity.
\begin{figure}[H]\label{fig2}
\centering
\includegraphics[width=0.5\textwidth]{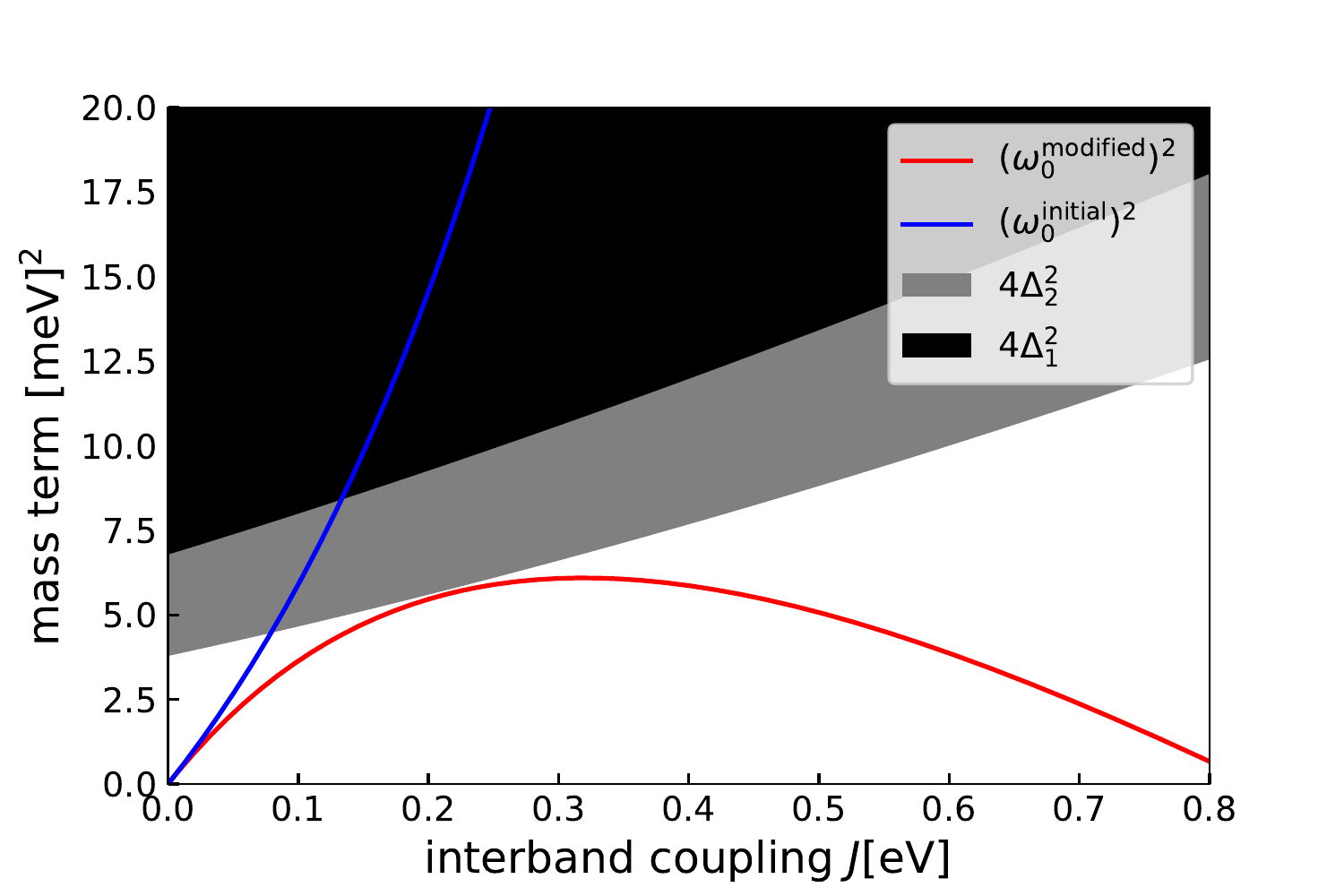}
\caption{Mass term of the Leggett mode when one changes the interband coupling with each coupling constant being $V_{11}=1.0[\mathrm{eV}][\mathrm{cell}], V_{22}=0.7[\mathrm{eV}][\mathrm{cell}], \rho_1=\rho_2=0.3[\mathrm{eV}]^{-1}[\mathrm{cell}]^{-1}$. Red line and blue line correspond to $\left(\omega_0^{\mathrm{modified}}\right)^2$ and $\left(\omega_0^{\mathrm{initial}}\right)^2$, respectively. Black and gray shaded areas mean particle-hole continuum for $\Delta_1, \Delta_2$}
\centering
\includegraphics[width=0.5\textwidth]{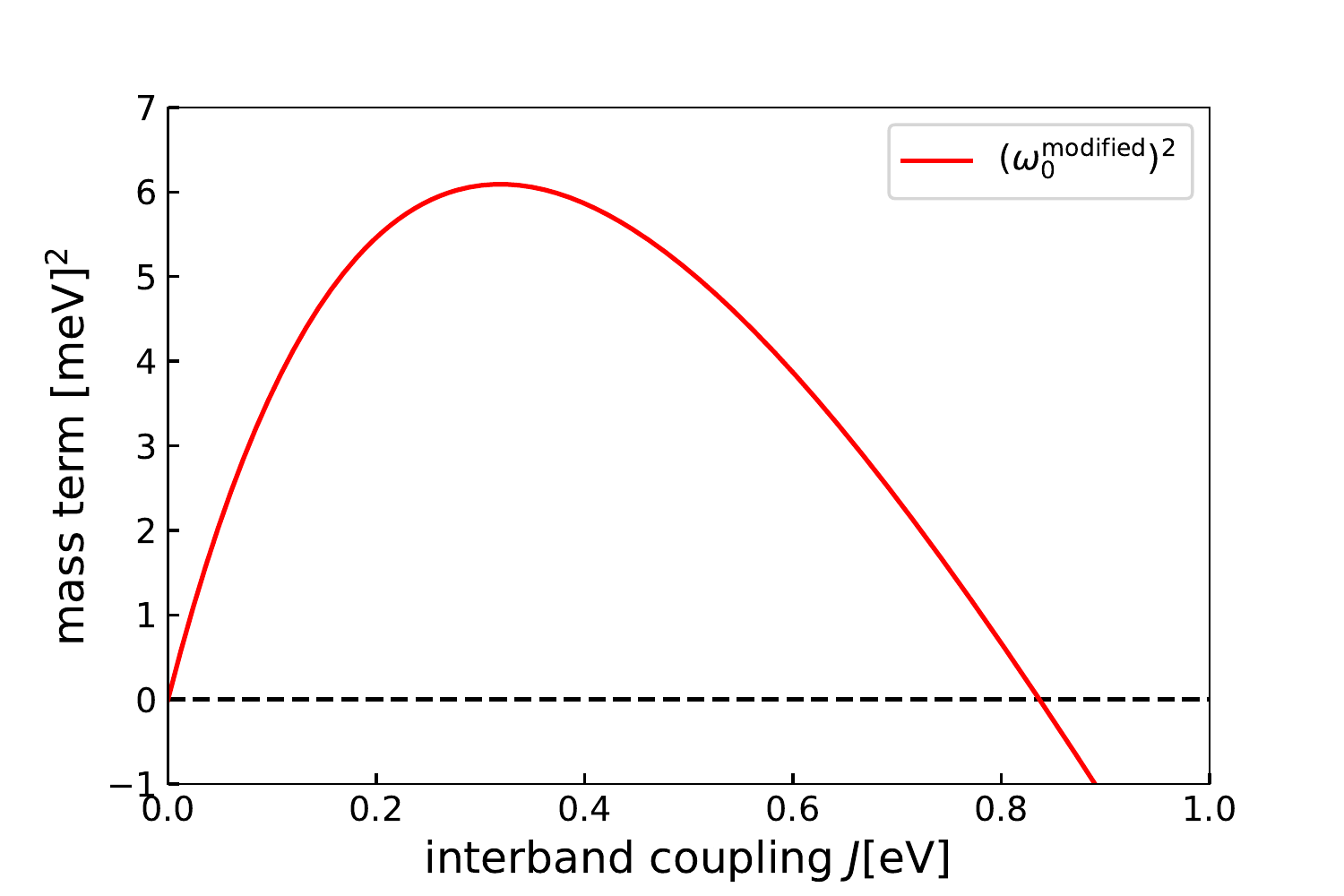}
\caption{Mass term of the Leggett mode when one changes the interband coupling with each coupling constant being $V_{11}=1.0[\mathrm{eV}][\mathrm{cell}], V_{22}=0.7[\mathrm{eV}][\mathrm{cell}], \rho_1=\rho_2=0.3[\mathrm{eV}]^{-1}[\mathrm{cell}]^{-1}$. When $J^2>V_{11}V_{22}$, $\left(\omega_0^{\mathrm{modified}}\right)^2$ becomes negative. }
\end{figure}
\subsection{Vanishing of the Leggett mode}
As we saw in \ref{222}, if the dimensionless parameter $\lambda_i\quad(i=1,2)$ is small enough, the result (\ref{14}) and (\ref{16}) match the earlier results (\ref{15}) or (\ref{18}) up to linear order in $J$. We can also see from (\ref{14}) that when higher terms in $\lambda_i$ cannot be neglected, (\ref{14}) and (\ref{16}) become substantially different from (\ref{15}) and (\ref{18}) as one can also see from Fig. 1.

Since experimentally the mass term of the Leggett mode, $\omega_0$ is usually probed by Raman spectroscopy, here mass term will be considered. One thing we can see from Sec. \ref{222}, or Fig. 1 is that whether the given system is neutral or charged, the modified theory which is presented in this paper always has a smaller mass term than the earlier theory which was presented by Sharapov et al.. This means phase twist defined by (\ref{7}) is more favorable.

One should realize by looking at (\ref{14}) that the Leggett mode exists when $\lambda_i$ satisfies
\begin{equation}\label{cond}
0<\lambda_i<\frac{1}{2}.
\end{equation} 
Otherwise, the Leggett mode has negative mass. Note that in the earlier theory, our assumption ensures that the criteria for the existence of the Leggett mode is 
\begin{equation}\label{criteria}
0<V_{11}V_{22}-J^2.
\end{equation}

We will see that (\ref{cond}) is equivalent to (\ref{criteria}). First, let us assume that $\Delta_1$ is real and positive. Our choice of coupling constants makes $\Delta_2$ real and positive as well\cite{doi:10.1143/PTP.36.1111}. Therefore the condition (\ref{cond}) is equivalent to
$0<2J\Phi_{\overline{i}}<\Delta_i$.
Together with the gap equation $\Delta_i=V_{ii}\Phi_i+J\Phi_{\overline{i}}$, one finds $J\Phi_{\overline{i}}<V_{ii}\Phi_i$. Therefore the following relation holds:
\begin{equation*}
\begin{split}
0<\left(V_{11}V_{22}-J^2\right)\Phi_1\Phi_2
\Leftrightarrow0<\frac{\lambda_1\lambda_2}{J^2}\left(V_{11}V_{22}-J^2\right)\Delta_1\Delta_2.
\end{split}
\end{equation*}
Since $\lambda_i>0$, this is equivalent to $0<V_{11}V_{22}-J^2$, which is (\ref{criteria}).

This can be understood as follows. If one defines a matrix as a set of coupling constants such as
\begin{equation}
\hat{V}=\begin{pmatrix} V_{11} & J \\ J & V_{22}\end{pmatrix},
\end{equation}
then (\ref{criteria}) is equivalent to $\mathrm{det}\hat{V}>0$. If $\mathrm{det}\hat{V}\leq0$, then one of Hubbard-Stratonovich fields cannot be defined since $\hat{V}$ has a zero or negative eigenvalue. Therefore only a Hubbard-Stratonovich field corresponding to the positive eigenvalue is defined. In other words the system has only one order parameter. One order parameter immediately manifests vanishing of the relative phase fluctuation. 

Note that if one includes the Higgs mode $h_i(x)$ such as $\Phi_i(x)=\left(\Phi_i+h_i(x)\right)e^{i\theta_i(x)}$, this conclusion is not changed at zero temperature since $h_i(x)$ does not couple to $\theta_i(x)$. Therefore the Leggett mode vanish at the point where $\mathrm{det}\hat{V}=0$. 

\subsection{Comparison with the earlier theory}
Both the earlier theory and the modified theory predict vanishing of the Leggett mode, however the way it does is much different. In the earlier theory, the mass term becomes divergent, which means the Leggett mode is arbitrarily rigid while the modified theory predicts that the Leggett mode is arbitrarily soft. Let us see why. We require the following relation,
\begin{equation}\label{ultraman}
\begin{pmatrix} \Delta_1(x) \\ \Delta_2(x)\end{pmatrix}=\hat{V}\begin{pmatrix} \Phi_1(x) \\ \Phi_2(x)\end{pmatrix}.
\end{equation} 
Therefore fluctuations of the fields should also satisfy
\begin{equation}\label{relationoffields}
\begin{pmatrix} \delta\Delta_1(x) \\ \delta\Delta_2(x)\end{pmatrix}=\hat{V}\begin{pmatrix} \delta\Phi_1(x) \\ \delta\Phi_2(x)\end{pmatrix}\Leftrightarrow\hat{V}^{-1}\begin{pmatrix} \delta\Delta_1(x) \\ \delta\Delta_2(x)\end{pmatrix}=\begin{pmatrix} \delta\Phi_1(x) \\ \delta\Phi_2(x)\end{pmatrix},
\end{equation} 
where $\delta\Delta_i(x), \delta\Phi_i(x)$ are fluctuations around the mean-field values and $\hat{V}^{-1}$ is an inverse matrix of $\hat{V}$. (\ref{relationoffields}) shows that if $\mathrm{det}\hat{V}$ is small, then $\delta\Phi_i(x)$ must be large even if $\delta\Delta_i(x)$ is small. The earlier theory assumes small fluctuation of gap $\delta\Delta_i(x)$. Therefore fluctuation of pair wavefunction is large and large fluctuation costs high energy, which results in the divergent mass term. 

On the other hand, in the modified theory we assume that $\delta\Phi_i(x)$ is small, which results in small $\delta\Delta_i(x)$ as well. One can verify $\omega_0=0$ when $\mathrm{det}\hat{V}=0$ by the following argument although it does not have any physical meaning to consider the case of $\mathrm{det}\hat{V}=0$ in the above-mentioned reason in Sec. 3.1. (\ref{ultraman}) tells that $\Phi_i(x)$ is not determined even if $\Delta_i(x)$ is given. This ``virtual symmetry" (in reality the system does not have this symmetry) makes the Leggett mode look massless (the Leggett mode is not present at this point).

\section{Spectrum in Raman scattering}
We have so far seen that dispersion of the Leggett mode can become substantially different when higher correction in interband coupling is taken into account. Experimentally, one can observe the signal of the Leggett mode in Raman spectroscopy. Peak corresponding to the Leggett mode captures the mass of the Leggett mode.

In Raman scattering, incident photons couple to density fluctuation such that
\begin{equation*}
\rho_R(\bm{p},\tau)=\sum_{i\sigma}\int\frac{d\bm{k}}{(2\pi)^d}\gamma_i(\bm{k})\psi_{i\sigma}^\dagger\left(\bm{k}-\frac{\bm{p}}{2},\tau\right)\psi_{i\sigma}\left(\bm{k}+\frac{\bm{p}}{2},\tau\right),
\end{equation*}
where the Raman vertex 
\begin{equation*}
\gamma_i(\bm{k})=m_i\sum_{\alpha,\beta}e_\alpha^I\frac{\partial^2\epsilon_i(\bm{k})}{\partial k_\alpha\partial k_\beta}e_\beta^S,
\end{equation*}
with band dispersion of $i$th band $\epsilon_i(\bm{k})$ and $e_{\alpha/\beta}^{I/S}$ being $\alpha/\beta$ component of the polarization of incident/scattering photon.

The response function $\chi_{RR}$ is 
\begin{equation*}
\chi_{RR}(\bm{p},\tau-\tau')=-\braket{T_\tau\rho_R(\bm{p},\tau)\rho_R(-\bm{p},\tau')},
\end{equation*}
where $T_\tau$ is time-ordered product in imaginary time $\tau$ and $\braket{\cdots}$ is a statistical average. 

$\chi_{RR}(\bm{p},\tau-\tau')$ can be expressed in frequency representation such that
\begin{equation*}
\chi_{RR}(\bm{p},i\nu_l)=\frac{1}{\sqrt{\beta}}\int_0^\beta d\tau\chi_{RR}(\bm{p},\tau-\tau')e^{i\nu_l(\tau-\tau')}.
\end{equation*}
By the linear response theory, spectral intensity is calculated as follows:
\begin{equation*}
S_R(\omega)=-\frac{1}{\pi}(1+n_B(\omega))\mathrm{Im}\chi_{RR}(\bm{p}=\bm{0},\omega),
\end{equation*}
where $n_B(\omega)$ is the Bose distribution function, i.e., $n_B(\omega)=\left(e^{\beta\omega}-1\right)^{-1}$ and $\chi_{RR}(\bm{p},\omega)$ is obtained by analytic continuation $i\nu_l\rightarrow\omega+i0$ from $\chi_{RR}(\bm{p},i\nu_l)$.

In order to calculate the response function, one can add the source term $S_J$ to the effective action:
\begin{equation*}
\begin{split}
S_J&=\int_0^\beta d\tau\int\frac{d\bm{p}}{(2\pi)^d}\rho_R(\bm{p},\tau)J(-\bm{p},\tau),\\
Z[J]&=\int\mathcal{D}\psi^\dagger\mathcal{D}\psi\mathcal{D}\Phi^*\mathcal{D}\Phi\mathcal{D}\varphi e^{-(S_{\mathrm{charged}}+S_J)}.
\end{split}
\end{equation*}
The response function is obtained by the functional differentiation
\begin{equation*}
\chi_{RR}(\bm{p},\tau-\tau')=-\left.\frac{1}{Z[0]}\frac{\delta^2Z[J]}{\delta J(-\bm{p},\tau)\delta J(\bm{p},\tau')}\right|_{J=0}.
\end{equation*}
In a charged superconductor, the response function in momentum representation is as follows:
\begin{equation}\label{r}
\begin{split}
&\chi_{RR}(\bm{p},i\nu_l)=\chi_{RR}^0(\bm{p},i\nu_l)-\begin{pmatrix}J_{\theta R}^1(-p) & J_{\theta R}^2(-p) & J_{\varphi R}(-p) \end{pmatrix}G_{\theta;\mathrm{charged}}\begin{pmatrix} J_{\theta R}^1(p) \\ J_{\theta R}^2(p) \\ J_{\varphi R}(p)\end{pmatrix}
\end{split},
\end{equation}
where 
\begin{equation*}
\begin{split}
\chi_{RR}^0(\bm{p},i\nu_l)&=\frac{1}{2\beta}\sum_n\int\frac{d\bm{k}}{(2\pi)^d}\sum_i\gamma_i\left(\bm{k}+\frac{\bm{p}}{2}\right)^2\pi_i^{33}(k,p)\\
J_{\theta R}^i(p)&=\frac{1}{2\beta}\sum_n\int\frac{d\bm{k}}{(2\pi)^d}\left[-\gamma_i\left(\bm{k}+\frac{\bm{p}}{2}\right)\left(\frac{\nu_l}{2}\pi^{33}_i(k,p)+\frac{i\hbar^2}{4m_i}\left(2\bm{k}+\bm{p}\right)\cdot\bm{p}\pi^{03}_i(k,p)+J\Phi_{\overline{i}}\pi^{23}_i(k,p)\right)\right.\\&\left.+\gamma_{\overline{i}}\left(\bm{k}+\frac{\bm{p}}{2}\right)J\Phi_i\pi_{\overline{i}}^{23}(k,p)\right]\\
J_{\varphi R}(p)&=\frac{-i}{2\beta}\sum_n\int\frac{d\bm{k}}{(2\pi)^d}\sum_i\gamma_i\left(\bm{k}+\frac{\bm{p}}{2}\right)\pi_i^{33}(k,p).
\end{split}
\end{equation*}
The first term in the right hand side of (\ref{r}) corresponds to the peaks which probe the gap $\Delta_i$ and the second term shows collective excitations. Since in Raman scattering it suffices to know the response function with zero momentum $\bm{p}=\bm{0}$, poles are located at the mass of the Leggett mode and plasma frequency. Therefore one can observe the peak corresponding to the Leggett mode $\omega_{\mathrm{peak}}$ at the mass of the Leggett mode, i.e., $\omega_{\mathrm{peak}}=\omega_0^{\mathrm{modified}}$ in this setting.

It is obvious to see that in a neutral superconductor the position of peak for the Leggett mode is located at $\omega_{\mathrm{peak}}=\omega_0^{\mathrm{modified}}$.

To compare how difference of theories affects experimental observation it suffice to see the difference of the mass term of the Leggett mode, i.e., $\omega_0^{\mathrm{initial}}$ and $\omega_0^{\mathrm{modified}}$.

Figure 5 shows that if $J=V_{12}$, i.e., interband coupling is small so that $\lambda_i\ll1$ holds, $\omega_0^{\mathrm{initial}}$ and $\omega_0^{\mathrm{modified}}$ are almost the same, but as the interband coupling is large they become much different. In this model each coupling constant is set as $V_{11}=1.0[\mathrm{eV}][\mathrm{cell}], V_{22}=0.7[\mathrm{eV}][\mathrm{cell}], \rho_1=\rho_2=0.3[\mathrm{eV}]^{-1}[\mathrm{cell}]^{-1}$. Note that as is usual with real experiments, generally $\omega_0^{\mathrm{initial}}$ is above quasiparticle continuum, i.e., $\omega_0^{\mathrm{initial}}>2\Delta_i (i=1,2)$ (Fig. 5). This makes experimental observation of the Leggett mode difficult because peak is broadened. On the other hand, Fig. 5 shows that for strong coupling $\omega_{\mathrm{peak}}$ is below quasiparticle continuum and therefore clearly detectable, i.e., $\omega_0^{\mathrm{modified}}<2\Delta_i$. Note that this result comes from the oversimplified two band model and we cannot conclude that in real materials such as \ce{MgB2} this is also true at this stage.

\begin{figure}
\centering
\includegraphics[width=0.5\textwidth]{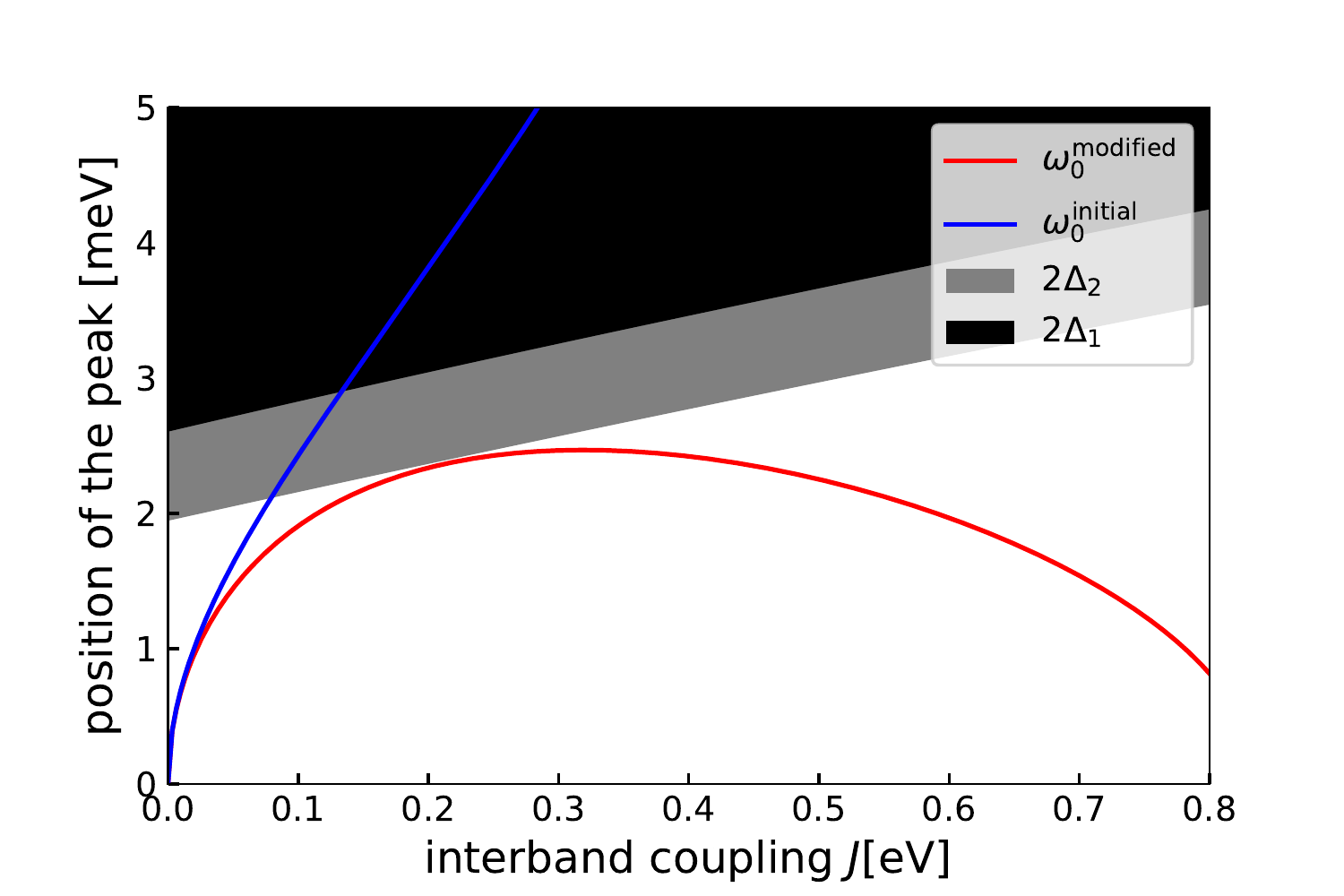}
\caption{Spectral peak $\omega_0^{\mathrm{modified}}$ and $\omega_0^{\mathrm{initial}}$ as functions of interband coupling $J=V_{12}$ with coupling constants $V_{11}=1.0[\mathrm{eV}][\mathrm{cell}], V_{22}=0.7[\mathrm{eV}][\mathrm{cell}], \rho_1=\rho_2=0.3[\mathrm{eV}]^{-1}[\mathrm{cell}]^{-1}$. Black and gray shaded area correspond to quasiparticle continuum for $\Delta_1$ and $\Delta_2$, respectively.}
\end{figure}

\begin{figure}
\centering
\includegraphics[width=0.47\textwidth]{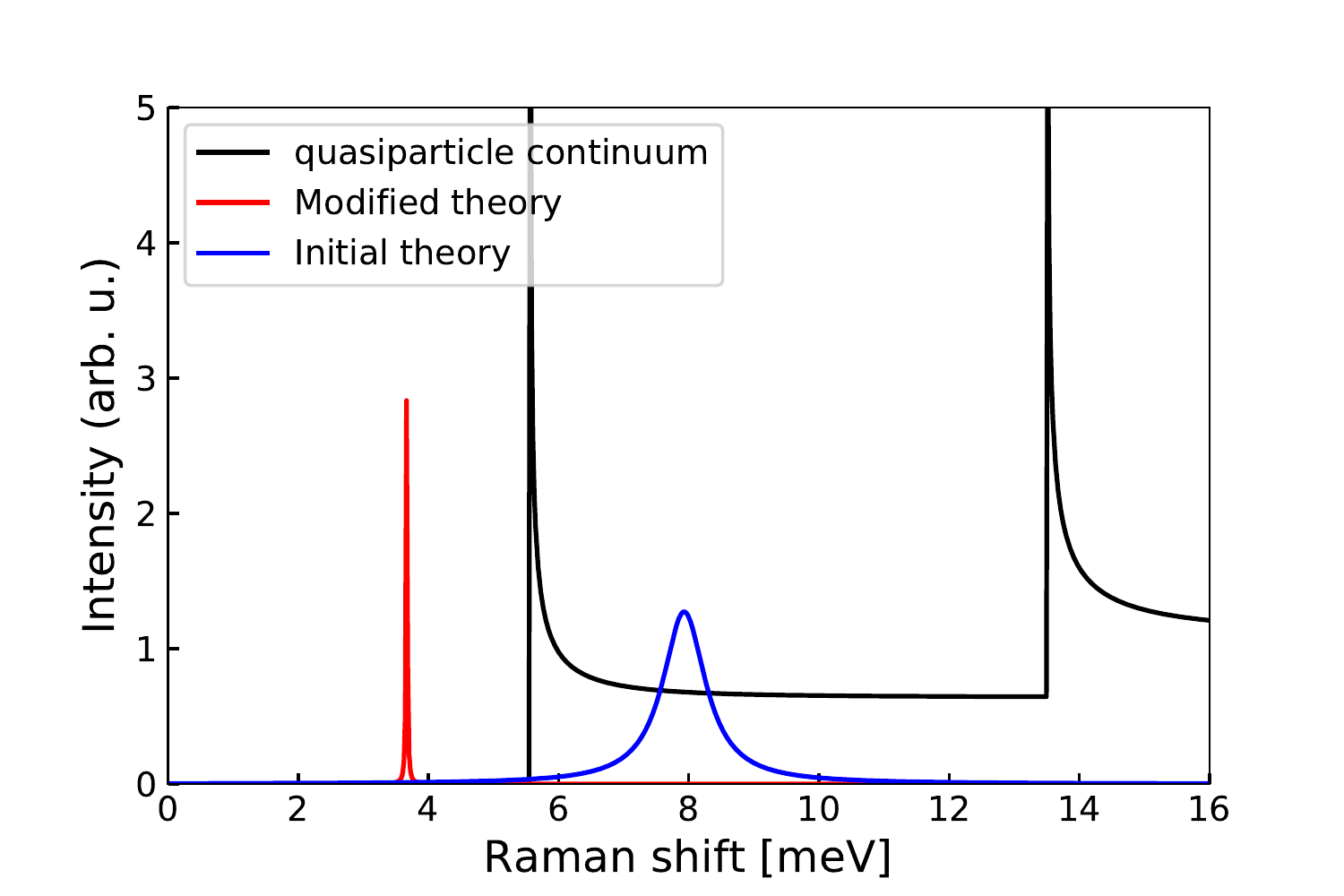}
\caption{Schematic view of spectral intensity. Black line corresponds to quasiparticle continuum $2\Delta_i (i=1,2)$ and colored lines to the  Leggett mode for the modified theory and derived from the initial theory, respectively. Due to Landau damping, peak become broadened if position of the peak is larger than $2\Delta_i (i=1,2)$.}
\end{figure}

\section{Conclusion}
We first looked at the validity of the effective action approach to describe collective excitations with two band model, and saw that the set of phase transformation defined in (\ref{7}), that is,
\begin{equation*}
\begin{split}
\psi_{i\sigma}(x)&\rightarrow e^{i\theta_i(x)/2}\psi_{i\sigma}(x)\\
\Delta_i(x)&\rightarrow e^{i\theta_i(x)}\Delta_i(x),
\end{split}
\end{equation*}
is not consistent with the definition of the gap. Instead, one can introduce the consistent set of phase fluctuation, which is
\begin{equation*}
\begin{split}
\psi_{i\sigma}(x)&\rightarrow e^{i\theta_i(x)/2}\psi_{i\sigma}(x)\\
\Phi_i(x)&\rightarrow e^{i\theta_i(x)}\Phi_i(x),
\end{split}
\end{equation*}
where $\Phi_i(x)$ is pair wavefunction. From this set of transformations dispersions for phase has been derived. Dispersions show the existence of the massless mode, BAG mode, and the massive mode, the Leggett mode in a neutral superconductor. If one compare dispersions derived in this paper with those in the earlier theory, they are the same if higher order in interband coupling can be ignored, on the other hand, those are much different with large interband coupling.

The mass term of the Leggett mode is sensitively dependent on interband coupling 
, and it has been shown that for both a neutral system and a charged system the mass term derived in this paper is always smaller than the mass derived in the earlier theory, which justifies the phase twist (\ref{7}). For a large interband coupling, only one Hubbard-Stratonovich field is defined so that the Leggett mode is no longer present.


\begin{acknowledgments}
This work is originally from a report of the semester project, one of the courses at ETH Z\"{u}rich where I spent one year as an exchange student. I thank my supervisor at ETH Z\"{u}rich, M. Sigrist for a number of his useful discussion and advices. I also thank Y. Kato who is my supervisor at University of Tokyo for his comments and discussions, and S. Sharapov for his discussion. 
\end{acknowledgments}

\bibliographystyle{unsrt}
\bibliography{jpsj_eth}

\end{document}